\documentclass[trackchanges]{aastex701}
\usepackage{graphicx} 
\usepackage{natbib} 
\usepackage{amssymb}
\usepackage{amsmath}
\usepackage{gensymb}
\usepackage{textcomp}
\usepackage{upgreek}
\usepackage{makecell}
\usepackage{color}
\usepackage{hyperref}

\newcommand\aastex{AAS\TeX}

\newcommand{\ycc}[1]{\textcolor{red}{[YC: #1]}}

\begin{document}
\title{Mock Observations of Multiple Stellar Populations in Tidal Streams of Palomar 5 for the Chinese Space Station Survey Telescope}

\author[orcid=0000-0001-6820-1683]{Xia Li}
\affiliation{School of Physics and Astronomy, Sun Yat-sen University, Daxue Road, Zhuhai, 519082, China}
\email[show]{lixia76@mail2.sysu.edu.cn}  

\author[orcid=0000-0001-8713-0366]{Long Wang} 
\affiliation{School of Physics and Astronomy, Sun Yat-sen University, Daxue Road, Zhuhai, 519082, China}
\affiliation{CSST Science Center for the Guangdong-Hong Kong-Macau Greater Bay Area, Zhuhai, 519082, China}
\email[show]{wanglong8@sysu.edu.cn}

\author{Chengyuan Li}
\affiliation{School of Physics and Astronomy, Sun Yat-sen University, Daxue Road, Zhuhai, 519082, China}
\affiliation{CSST Science Center for the Guangdong-Hong Kong-Macau Greater Bay Area, Zhuhai, 519082, China}
\email{lichengy5@mail.sysu.edu.cn}

\author{Yang Chen}
\affiliation{School of Physics and Optoelectronic Engineering, Anhui University, Hefei, 230601, People’s  Republic of China}
\email{cy@ahu.edu.cn}

\author{Hao Tian}
\affiliation{National Astronomical Observatories, Chinese Academy of Sciences, Beijing 100101, People's Republic of China}
\email{tianhao@nao.cas.cn}

\author{Xin Zhang}
\affiliation{Key Laboratory of Space Astronomy and Technology, National Astronomical Observatories, Chinese Academy of Sciences, 20A Datun Road, Beijing 100101,People’s Republic of China}
\email{zhangx@bao.ac.cn}

\begin{abstract}

Observations show that multiple stellar populations (MPs) are ubiquitous in globular clusters. The Hubble Space Telescope (HST) has been a pivotal tool for previous photometric studies of MPs. The Chinese Space Station Survey Telescope (CSST) is a two-meter telescope scheduled for launch. One of its imaging instruments, the Survey Camera (SC), combines ultraviolet sensitivity comparable to that of HST with a significantly larger field of view, making it well-suited for conducting large-scale photometric surveys of MPs within extensive stellar stream structures. In this work, we perform mock observations of the stellar stream Palomar 5 to assess the feasibility of detecting MPs with the CSST/SC. The results indicate that the CSST/SC cannot resolve MPs in stellar streams at distances comparable to Palomar 5 ($\gtrsim 20$ kpc) with one or ten 150 s exposures. This fundamental limitation arises from the absence of the precise proper motions required to disentangle stream members. We estimate that successful resolution would require the target stream to be $\lesssim$ 8 kpc under a 150 s exposure. Furthermore, using theoretical color-magnitude diagrams, we find that the CSST/SC $g$-band provides an optimal balance between contamination rate and completeness rate for member identification in the cluster's core. However, this approach fails in the stream due to severe field star contamination. Therefore, future CSST observations of Palomar 5 and its tidal tails will employ multiple epochs across several bands to obtain the deep photometry and proper motion data for a definitive MP analysis.

\end{abstract}

\keywords{\uat{Globular clusters}{656} --- \uat{Stellar abundances}{1577} --- \uat{Photometry}{1234} --- \uat{Space telescopes}{1547}}

\section{Introduction}
\label{1}

Globular clusters (GCs) were traditionally considered to host a single stellar population \citep{Kraft1979}. However, advances in high-precision spectroscopy and photometry have revealed that most well-studied GCs ($\gtrsim$ 2 Gyr) in fact host multiple stellar populations (MPs) \citep{2022Univ....8..359M}. Particularly thanks to the high resolution Hubble Space Telescope (HST) ultraviolet (UV)-optical photometric observations, research on MPs has made significant progress \citep{2015ApJ...808...51M,2015AJ....149...91P, 2015MNRAS.447..927M, 2017MNRAS.464.3636M}.

These MPs are characterized by the following features: (1) a first population (1P) with chemical abundances similar to those of field stars of comparable metallicity; one or more second populations (2P), typically enriched in He, N, Na, and Al but depleted in C, O, and in some cases Mg \citep{2011A&A...531A..35P, 2014ApJ...786...14M, 2014ApJ...795L..28C, 2014ApJS..210...10R, 2015ApJ...808...51M, 2017MNRAS.464.3636M}; (2) distinct variations in light elements among the 2P stars, evident as the well-known C-N, O-Na, and Mg-Al anticorrelations \citep{2005AJ....129..303C, 2009A&A...505..139C, 2019AJ....158...14N}; (3) a generally constant sum of C+N+O abundances \citep{2005AJ....129..303C, 2016MNRAS.459..610M}; and (4) in rare cases, variations in heavier elements, including $\alpha$-elements (Si, Ca, Ti), iron-peak elements (Fe, Ni), and neutron-capture elements (Sr, Ba, La, Eu) \citep{2011A&A...532A...8M, 2016MNRAS.455.2417R}. Moreover, the presence of MPs appears to be a universal phenomenon, observed in stars at various evolutionary stages from the main sequence (MS) to the giant phase \citep{2012ApJ...744...58M, 2014AJ....148...27C, 2017A&A...603A..37G}. Furthermore, as stars in the MS and red giant branch (RGB) phases are largely unaffected by significant mass loss from stellar winds, they can serve as reliable probes of the initial abundances that produce the MP features seen in color-magnitude diagrams (CMDs). 

The ubiquity of these chemical anomalies across all stellar evolutionary stages in several GCs rules out internal mixing within the star as the explanation for MPs \citep{grattonAbundanceVariationsGlobular2004}. Most models instead propose that 2P stars formed from material ejected by 1P polluters. These models can be broadly divided into multi-generation scenarios, in which 2P stars form from the enriched ejecta of earlier polluters in subsequent star-formation episodes \citep{2022Univ....8..359M}. Multi-generation scenarios typically involve polluters such as intermediate-mass asymptotic giant branch stars (AGB; 3-8 $M_{\odot}$) \citep{10.1111/j.1365-2966.2010.16996.x, 2016MNRAS.458.2122D}, fast-rotating massive stars (FRMSs; $>$ 15 $M_{\odot}$) \citep{2007A&A...464.1029D, 2013A&A...552A.121K}, massive interacting binaries \citep{2009A&A...507L...1D}, super massive stars ($>$ 1,000 $M_{\odot}$) \citep{2014MNRAS.437L..21D}. Alternative scenarios propose that all GC stars are contemporaneous, suggesting that the distinctive chemical signatures of the 2P stem from accretion processes occurring during the pre-MS phase \citep{2013MNRAS.436.2398B, 2018MNRAS.478.2461G}. These scenarios include early disc accretion from massive interacting binaries and FRMSs \citep{2013MNRAS.436.2398B}, super massive stars (super massive stars formed via runaway collisions) \citep{2018MNRAS.478.2461G}, mergers of massive stars ($>$ 30 $M_{\odot}$) \citep{2020MNRAS.491..440W} in dense clusters, and other non-canonical channels operating under regulated or delayed feedback \citep{2013A&A...552A.121K, 2022MNRAS.513.2111R}.

A key clue to the formation history of MPs is the spatial distribution of stars within each population \citep{2023MNRAS.520.1456L}. Studies have shown that in most GCs, 2P stars are more centrally concentrated than 1P stars \citep{2022Univ....8..359M, 2023A&A...677A...8O, 2024A&A...681A..45L}. Theoretical models of MPs can be invoked to explain this central concentration of 2P stars. However, exceptions exist: in some clusters, 1P stars are more centrally concentrated than 2P stars \citep{2023MNRAS.520.1456L}, or both populations share identical radial distributions \citep{2015MNRAS.454.2166M}. Therefore, investigating the spatial distribution of MPs offers crucial constraints on their formation mechanisms and may provide insights for new theoretical explanations.

 However, long-term cluster evolution drives strong dynamical mixing that can erase initial spatial information, whereas in low-density tidal tails formed by escaping stars, mixing is significantly slower. Moreover, different regions along the streams preserve information about the cluster's initial mass function (IMF), MP ratios, and binary fraction. Recovering such evolutionary information from tidal streams would help reconstruct the cluster's initial properties. Therefore, studying the spatial distribution of different stellar populations in tidal streams offers novel insights into the formation mechanisms of MPs. Recent studies have reported MPs in several GC streams, such as 300S \citep{2024MNRAS.529.2413U}, GD-1 \citep{2022MNRAS.515.5802B}, and Phoenix \citep{wan_tidal_2020, 2021ApJ...921...67C}. Although the number of stream stars with confirmed MP signatures is still limited, high-resolution spectroscopic observations have consistently revealed significant dispersions in their chemical abundances.

Observations of MPs typically rely on spectroscopic and photometric methods. High-resolution spectroscopy can reveal variations in light element abundances through specific spectral lines, while photometry reveals distinct stellar sequences in various passbands. Helium, however, is difficult to measure directly from spectra due to its spectral line sensitivity to stellar temperature. Furthermore, obtaining high signal-to-noise ratio (SNR) and high-resolution spectra for large samples is observationally expensive. In contrast, photometry can detect helium variations \citep{2007ApJ...661L..53P, 2015MNRAS.446.1672M, 2018MNRAS.481.5098M} and proves particularly effective for identifying MPs  by using CMDs or pseudo-color-color diagrams (``chromosome maps'') within the crowded areas of star clusters \citep{2008A&A...490..625M, 2015ApJ...808...51M,2015AJ....149...91P, 2015MNRAS.447..927M,2017MNRAS.464.3636M}. 

The Chinese Space Station Survey Telescope (CSST) is a two-meter telescope with an off-axis three-mirror anastigmat \citep{Zhan2021}, which is scheduled for launch. It will host two primary imaging instruments: the Survey Camera (SC) and the Multi-Channel Imager (MCI). The CSST/SC will provide high spatial resolution ($\sim$0.15$''$) imaging across a wide wavelength range from the near-ultraviolet (NUV) to the near-infrared (NIR). With a wavelength coverage and spatial resolution comparable to HST but a significantly larger field of view (approximately 1.1$\degree$ × 1.0$\degree$), the CSST is well-suited for observing large-scale stellar streams. Furthermore, it has been demonstrated that both the SC and MCI can resolve MPs in star clusters using their UV capabilities  \citep{ji2023study,2022RAA....22i5004L}.

Leveraging its broad wavelength coverage and wide field of view (FoV), the CSST/SC instrument is expected to greatly advance research in star cluster physics. Furthermore, by combining evolutionary information from stellar streams and high-precision \textsc{petar} simulations, we can develop accurate evolutionary models of GCs to investigate the spatial distribution of MPs. In this paper, we will focus on conducting a pre-operative study on CSST/SC's capability to observe MPs in stellar streams before its formal commissioning. Subsequent work will explore how CSST/SC-detectable stellar populations in streams relate to the cluster’s initial conditions under varying spatial distributions. 

One of the most interesting stellar streams to study is the Galactic halo GC Palomar 5. It possesses well-characterized tidal streams extending over $\sim 30^\circ$ on the sky \citep{odenkirchenDetectionMassiveTidal2001, odenkirchen2003extendeda, ibata2017feeling, bonacaVariationsWidthDensity2020}. The Palomar 5 cluster center is located at celestial coordinates ($\alpha$, $\delta$) $\approx$ ($229^\circ$, $-0.11^\circ$) and at a distance of $\gtrsim$ 20 kpc \citep{starkman2020extended}. It is metal poor and of low central density, which suggests that there may be a large number of black holes (BHs) in this cluster \citep{gieles2021supramassive, wang2023influence}. 

By integrating N-body simulation data of Palomar 5, we will generate isochrones with varying Helium abundance. These models will then be used to evaluate the ability of the CSST/SC photometric system to detect MPs in stellar streams. It should be noted that previous studies by \citet{2022RAA....22i5004L} and \citet{ji2023study} have not yet taken into account the impact of background contamination, photometric noise and point spread function (PSF) on the CSST's ability to identify MPs during actual telescope observations. These factors will be considered in this work. The article is organized as follows: in Section \ref{2}, we introduce the details of our data and method. We show our main results in Section \ref{3}. A brief discussion of our results is presented in Section \ref{4}. Section \ref{5} concludes this work.

\section{Data and Method}
\label{2}

Building upon this framework, we will employ N-body simulation data from \textsc{petar} \citep{wang2020PETAR} to model the observational properties of Palomar 5. To assess the impact of MPs within the CSST/SC photometric system, we will integrate the simulation data with Dartmouth isochrones \citep{dotterDartmouthStellarEvolution2008}, which support variations in He abundance. We focus on two distinct populations: 1P characterized by a standard metal mixture with solar-scale compositions and a normal helium fraction, and 2P characterized by depleted C and O, and enriched N and He. For each population, we will synthesize spectra using the \textsc{spectrum} code and compute their photometric properties with the \textsc{ybc} code \citep{2019A&A...632A.105C}. Mock observations will then be generated using the CSST/SC simulation software. Finally, we will perform photometry on each image with Photutils code \citep{larry_bradley_2024_13989456} to produce a comprehensive mock star catalog of Palomar 5.

\subsection{N-body Simulation Data}
\label{2.1}

Our simulation data are derived from the previous work on Palomar 5 by \citet{wang2023influence}. The authors conducted simulations of Palomar 5-like clusters using the high-performance $N$-body code \textsc{petar} \citep{wang2020PETAR}. This code can efficiently model the evolution of star clusters and tidal streams. It uses the high-performance parallel framework \textsc{fdps} \citep{Iwasawa2016,Iwasawa2020}, which implements the hybrid $P^{3}T$ method \citep{Oshino2011}: a high-precision direct $N$-body integrator (4th-order Hermite with slowdown-algorithmic regularization (SDAR) method \citep{2020MNRAS.493.3398W}) for short-range interactions and multiple systems, coupled with a particle-tree method \citep{Barnes1986} for long-range forces. 
The code also includes updated stellar evolution codes (\textsc{sse}/\textsc{bse} \citep{hurley2000comprehensive, hurley2002evolution, banerjee2020bse}) for evolving individual stars and binaries, and supports external galaxy potential through \textsc{galpy} \citep{bovy2015galpy}. These features make \textsc{petar} essential  for modelling star-by-star dynamical evolution of GCs with tidal tails.

\citet{wang2023influence} study how binaries and BHs affect the long-term evolution of Palomar~5. They use \textsc{petar} to build five star-by-star models with and without primordial binaries and BHs. These models can reproduce the observed surface number density of Palomar~5 and include tidal tail evolution under the MilkyWay potential MWPotential2014 from \textsc{galpy}.
Here, we select their noBin-BH model, which assumes many BHs retained in the cluster and no primordial binaries. The model starts with a half-mass radius of 5.85 pc and 2.1 $\times$ 10$^{5}$ stars, evolves to 11.5 Gyr at metallicity z = 0.0006, and uses a  \citet{1911MNRAS..71..460P} profile with the \citet{kroupa2001variation} IMF spanning $0.1$-$100$ $M_{\odot}$. 

During data processing, we removed core-helium-burning stars, white dwarfs, neutron stars, and BHs, as these stellar types are unavailable from the isochrones used later. Each simulated object has a unique identifiler ``id'' to track its evolutionary history, along with recorded properties: stellar mass, sky coordinates (RA and Dec), distance, and stellar type. 

The \textsc{sse} and \textsc{bse} stellar-evolution modules in \textsc{petar} do not support varying helium abundances or evolving MP, so the simulation data cannot directly reveal the MP distribution.
We therefore post-process the data: select subsets of stars as 1P and 2P and use isochrones with varying abundances to generate the mock MP photometry. The following sections describe this procedure in detail.

\subsection{Stellar Isochrones with Different Helium Abundances}
\label{2.2}

We used the Dartmouth Stellar Evolution Database \citep{dotterDartmouthStellarEvolution2008} to create two isochrones with different He abundances: one for 1P (Y=0.2461) and another for 2P (Y=0.3300). To prevent changes in [Fe/H], we kept [Fe/H] fixed at the observed value \citep[-1.41;][]{ibataChartingGalacticAcceleration2024} for both 1P and 2P. Similarly to the work by \citet{liSearchingMultiplePopulations2022}, the [$\alpha$/Fe] is 0.0 dex. We used the same age (11.5 Gyr) for both stellar populations.

\begin{figure}
    \centering    
    \includegraphics[width=0.8\textwidth]{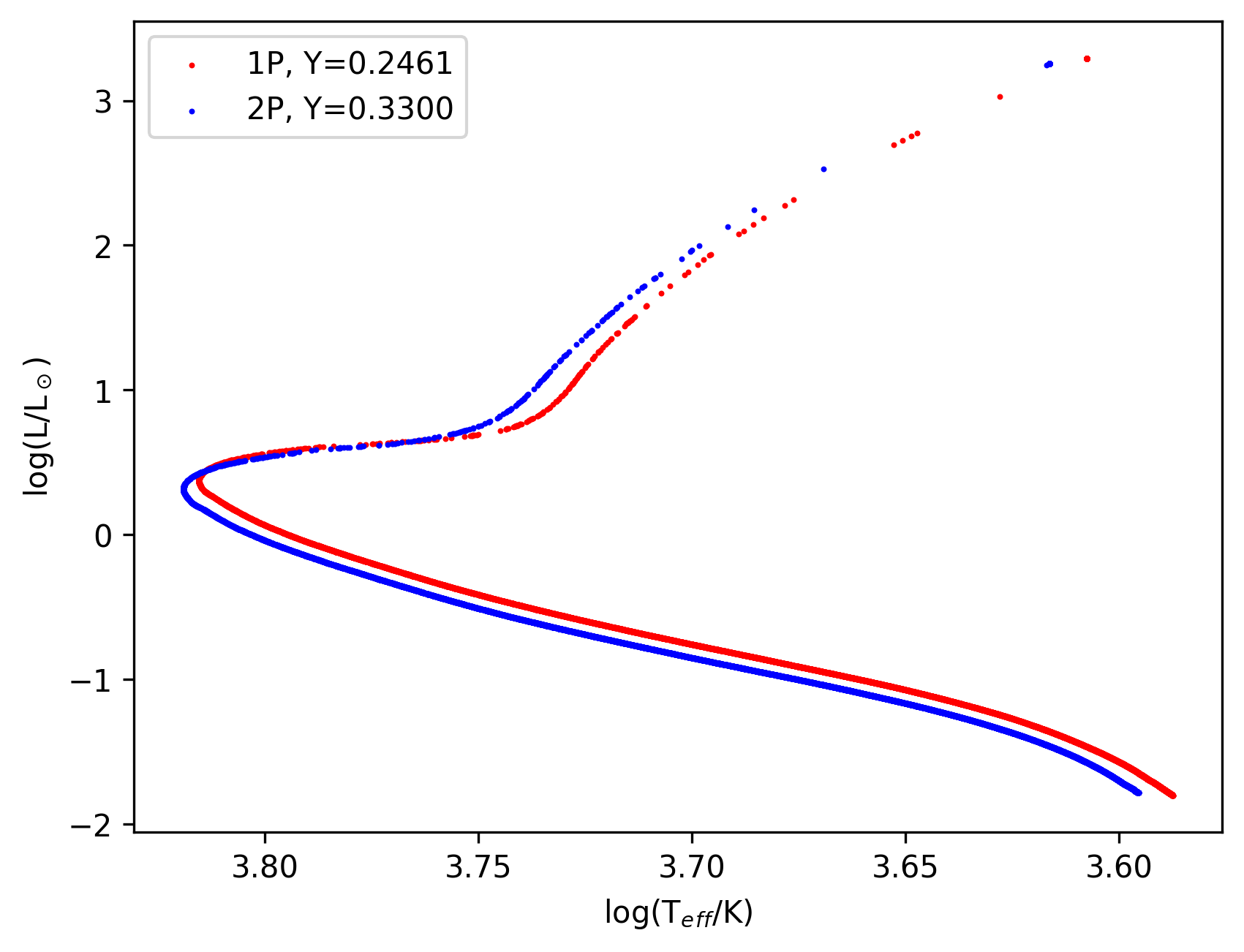}
    \caption{Dartmouth isochrones for Palomar 5 are displayed in the plane of $\log{(L/L_\odot})$ versus $\log{(T_{\mathrm{eff}})}$. The isochrones were linearly interpolated at [Fe/H] = –1.41 and ages from 11.5 Gyr. Red and blue dots denote 1P ($Y=0.2461$) and 2P ($Y=0.3300$) stars, respectively.}
    \label{fig:Figure1}
\end{figure}

The output isochrone contained nearly 200 data points and provided  stellar mass, effective temperature ($T_{\mathrm{eff}}$), surface gravity ($\log{g}$ in cgs units), and luminosity ($L$). For each stellar mass in the simulation data, we derived the corresponding 1P and 2P stellar parameters through linear interpolation of the output isochrones. In Figure \ref{fig:Figure1}, we show the isochrones in the Hertzsprung-Russell diagram (HR diagram). Consistent with the findings of \citet{ji2023study}, helium-enriched stars evolve more rapidly, becoming brighter and hotter at each evolutionary stage compared to normal stars. Finally, to examine the correlation between the formation mechanism  and the spatial distribution of different stellar populations, we assume a uniform and random distribution of 1P and 2P stars within Palomar 5. We assumed a 1:1 mixing ratio between the 1P and 2P populations. A mixing member star isochrone was then constructed through random sampling based on this ratio.

\subsection{Synthetic Spectra}
\label{2.3}

The absolute flux spectrum serves two key purposes: (1)  to serve as input for the CSST/SC mock observation pipeline, and (2) to derive  both absolute and apparent magnitudes. We used the package \textsc{spectrum} (version 2.77c) to synthesize the spectra. \textsc{spectrum} performs a standard computation assuming local thermodynamic equilibrium (LTE) and a plane-parallel atmosphere. We used \textsc{atlas9} \citep{castelliNewGridsATLAS92004} as input stellar atmosphere models. These models are based on the solar abundances by \citet{1998SSRv...85..161G}. The model grids are computed for $T_{\mathrm{eff}}$ from 3,500 K to 50,000 K, $\log{g}$ from 0.0 dex to 5.0 dex, and [M/H] = +0.5, +0.2, 0.0, -0.5, -1.0, -1.5, -2.0, -2.5, -3.5, -4, and -5.5. The input line lists were taken from the \textsc{spectrum} package website, and cover a wavelength range of 900 \AA–40000 \AA. According to the wavelength range of the CSST, we computed the synthetic spectra from 2500 to 11000 \AA.

We generated synthetic spectra both with 1P abundances (primordial abundances: a standard metal mixture with solar-scaled compositions) and enhanced in one of the light element abundances. Specifically, under the adoption of [Fe/H] = -1.41 and assuming that the total abundance of CNO remains unchanged, we set all 2P stars to have $\delta$Y = +0.0839, $\delta$[C/Fe] = -0.25, $\delta$[N/Fe] = +0.5, $\delta$[O/Fe] = -0.25 with respect to 1P stars through visual inspection based on rough estimation of Figure 4 and Figure 7 of \citet{phillips2022apogee}.

\subsection{YBC Bolometric Correction Database}
\label{2.4}

\begin{figure}
    \centering    
    \includegraphics[width=0.9\textwidth]{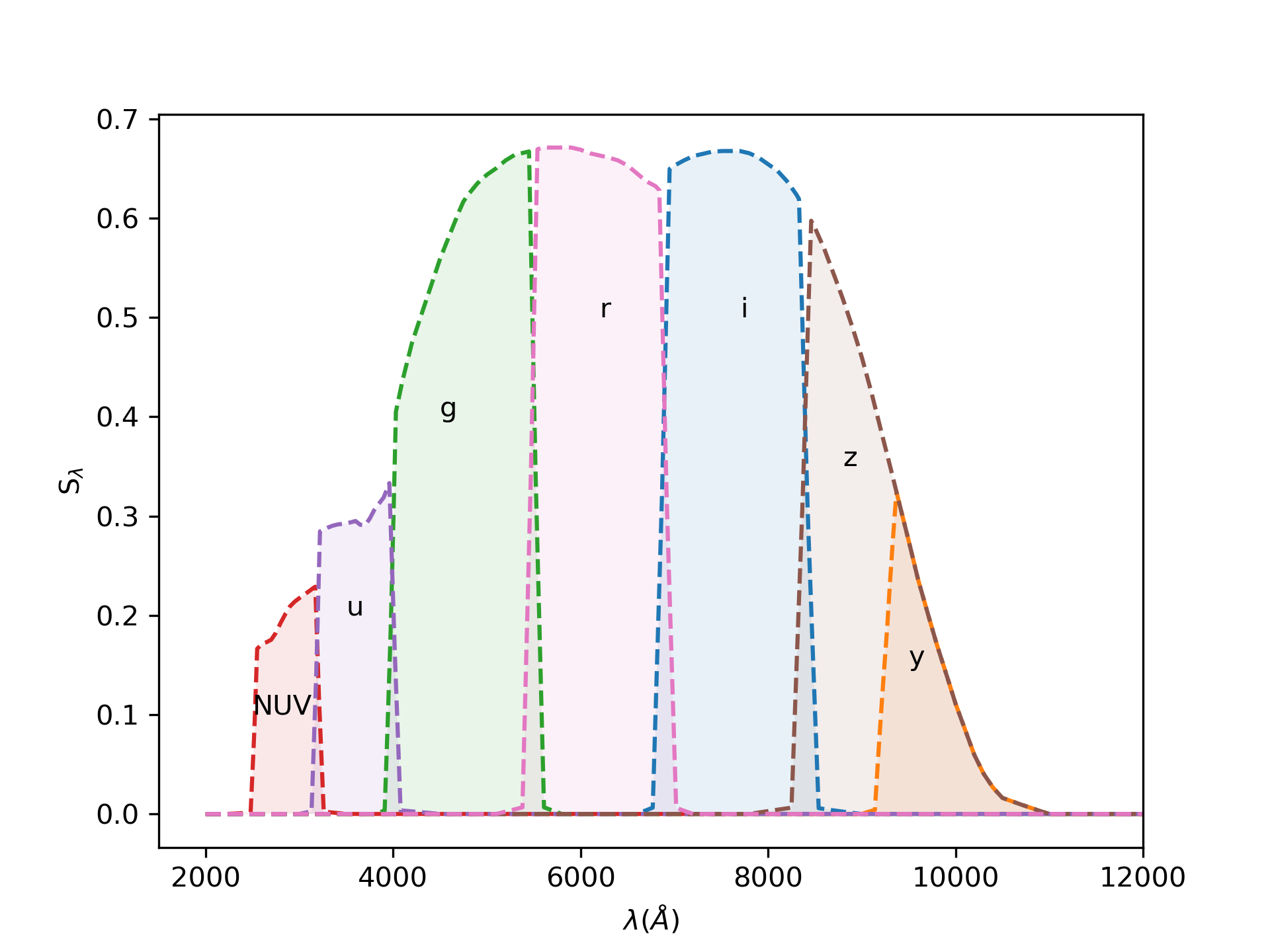}
    \caption{The total transmission curves for CSST/SC filter bands \citep{2025arXiv251106970W}.}
    \label{fig:Figure2}
\end{figure}

The CSST/SC is equipped with a seven-filter system, which includes the \textit{NUV}, $u$, $g$, $r$, $i$, $z$, and $y$ passbands (filter parameters \citep{collaboration2025introduction} are listed in Table \ref{tab:1}). The transmission
curves for CSST/SC filters \citep{2025arXiv251106970W} are presented in Figure \ref{fig:Figure2}. Then, we converted the model isochrones into observable CSST/SC magnitudes by \textsc{ybc} bolometric corrections database \citep{2019A&A...632A.105C}. The \textsc{ybc} database provides absolute magnitudes by combining stellar absolute flux spectrum, filter transmission curve, and stellar physical parameters ($T_{\mathrm{eff}}$, $\log{g}$, metallicity, extinction, etc.). We assumed an interstellar extinction of 0. Finally, using the distance to individual stars provided by the simulation data, we derived the apparent magnitudes for each star. These apparent magnitudes will serve as one of the parameters of the input star catalogs and as a theoretical apparent magnitude.

\begin{table}
    \centering
    \caption{Key parameters of the CSST/SC filters \citep{collaboration2025introduction}.}
    \begin{tabular}{cccc}
    \hline
    \hline
        Filter & Wavelength Range/nm & Exposure Time/s & \makecell{Limiting Magnitude  (Point source,\\ 5$\sigma$, AB magnitude systems)/mag} \\
        \hline
        NUV & 252-321 & 150 s $\times$ 4 & 25.4 \\
          u & 321-401 & 150 s $\times$ 2 & 25.4 \\
          g & 401-547 & 150 s $\times$ 2 & 26.3 \\
          r & 547-692 & 150 s $\times$ 2 & 26.0 \\
          i & 692-842 & 150 s $\times$ 2 & 25.9 \\
          z & 842-1080 & 150 s $\times$ 2 & 25.2 \\
          y & 927-1100 & 150 s $\times$ 4 & 24.4 \\
    \hline
    \end{tabular}
    \label{tab:1}
\end{table}

\subsection{CSST/SC Mock Observations}
\label{2.5}

In preparation for CSST operations, the CSST Science Ground Segment has developed a \textsc{galsim}-based \citep{ROWE2015121} software suite to generate simulated images. This tool \citep{2025arXiv251106970W} produces realistic mock observations suitable for scientific validation and is publicly available, supporting pixel-level CSST exposures at various fidelity levels. 

\subsubsection{Simulation Workflow}
\label{2.5.1}

The simulation workflow consists of the following stages: 

(1) Input star catalogs, survey strategies, absolute flux spectrum, and PSF samples are prepared separately from the imaging simulation. In particular, to produce a comprehensive set of realistic PSFs, an optical emulator was developed to simulate high-fidelity CSST/SC PSFs. It consists of six modules that model aberrations from mirror surface roughness, fabrication errors, CCD assembly errors, gravitational effects, and thermal distortions, as well as dynamic errors from micro-vibrations and image stabilization. 

(2) A spatially-varying, quasi-chromatic PSF model is then generated by interpolating the PSF samples in \textsc{galsim}. For each object, photon flux is assigned based on its magnitude and absolute flux spectrum through the corresponding filter, while its location on the image is determined by projecting celestial coordinates onto the pixel plane using the world coordinate system. 

(3) For each filter, the surface brightness profiles of objects are convolved with the local PSF model at their respective positions. The rendering is performed using the ``photon-shooting'' method in \textsc{galsim}. The final image is generated by stacking the PSF contributions from all sub-bandpasses and incorporating the simulated CSST background value along with various detector effects.

In general, this pipeline performs a chip-to-chip simulation. The incorporable physical effects include weak lensing shear, cosmic rays, and sky background, among others. Instrumental effects encompass PSF, readout noise, flat-fielding, shutter effects, dark current, charge transfer inefficiency (CTI), bad columns, etc. Users can independently run customized simulations according to their specific requirements. For this study, we utilize single epoch observations comprising a 150 s exposure. We will address the full depth of the wide survey and the deep field areas forecasts in forthcoming analyses.

\subsubsection{Input Star Catalogs}
\label{2.5.2}

\begin{figure}
    \centering    
    \includegraphics[width=0.9\textwidth]{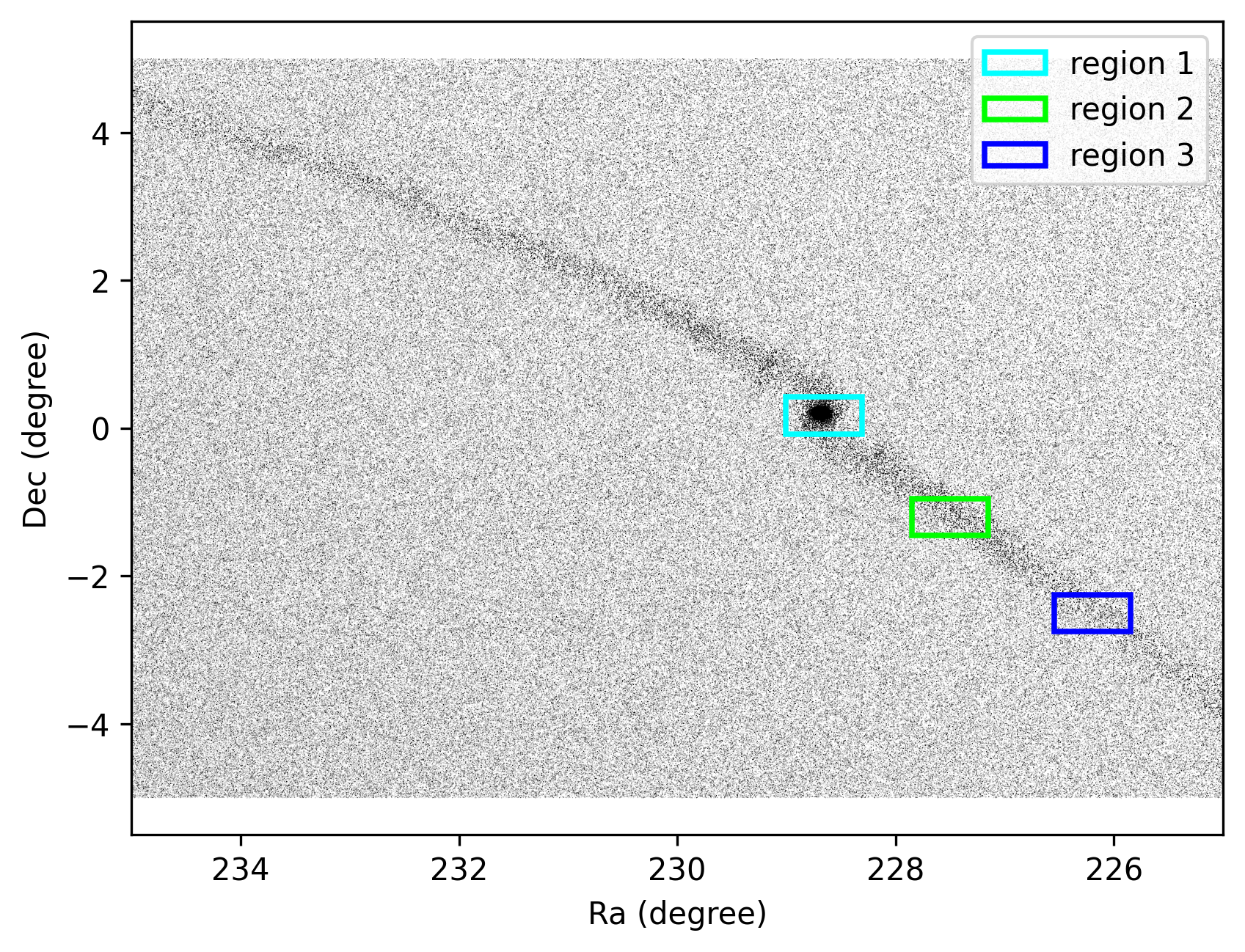}
    \caption{Spatial distribution of the entire sky region of Palomar 5 derived with CSST/SC g limiting magnitude. The boxes in different colors represent the final selected simulation regions, each with an identical area of $0.7^\circ$ $\times$ $0.5^\circ$. The cyan, green, and blue boxes correspond to the core region of the Palomar 5 cluster, the high-density segment, and the low-density segment of the tidal tail, respectively.}
    \label{fig:Figure3}
\end{figure}

In order to evaluate possible contamination of field stars, we used Galaxia code \citep{2011ApJ...730....3S} to create a field star catalog covering RA approximately from -$5^\circ$ to $5^\circ$ and Dec from 225° to 235°. For subsequent analysis, we labeled field stars and member stars (1P and 2P) in the input catalogs. Similarly, we simulated spectra using \textsc{spectrum} and derived their apparent magnitude by \textsc{ybc} database across the full metallicity range of field stars. It should be noted that in the process of synthesizing the spectrum, the \textsc{atlas9} stellar atmosphere models were used for both field stars and member stars. These atmosphere models impose certain constraints on their $T_{\mathrm{eff}}$, $\log{g}$, and [M/H]. Finally, the input catalog comprises both member stars and field stars, including star labels (1P/2P/field stars), the stellar mass, $T_{\mathrm{eff}}$, $\log{g}$, $L$, apparent magnitudes, RA, Dec, and distance, etc. We limit the input mock catalog to sources with CSST/SC $g$-band limiting magnitude and show its spatial distribution in Figure \ref{fig:Figure3}. The tidal tails of Palomar 5 are clearly visible above field star contamination across a RA range of approximately $225^\circ$ to $235^\circ$ (Figure \ref{fig:Figure3}). This is consistent with the spatial distribution map obtained from observations by \citet{bonacaVariationsWidthDensity2020}(Figure 2 in their paper). 

We selected three representative regions from the Palomar 5 cluster and its tidal structure with their field stars for mock observations. As shown in Figure \ref{fig:Figure3}, region 1, marked in cyan, represents the area of Palomar 5 cluster center. Region 2, indicated in green, corresponds to a high-density region in the tidal tail of Palomar 5. Region 3, highlighted in blue, denotes a low-density region of the tidal tail. In this paper, the dimensions of all three mock regions were set to $0.7^\circ$ $\times$ $0.5^\circ$.

\subsection{Data Reduction}
\label{2.6}

We used the Python code \textsc{photutils} \citep{larry_bradley_2024_13989456} for detecting and performing PSF photometry. The astrometry was anchored to an absolute reference frame using the CSST/SC mock observation output catalog, while the instrumental magnitudes were calibrated to the AB photometric system. To obtain high-precision photometry, we selected relatively isolated stars that were well fit by the PSF model and had small photometric and astrometric uncertainties to serve as photometric zero points in each image.

A preliminary catalog was compiled from all simulated photometric outputs with a SNR \textgreater\ 5 within the linear range. Then we removed overlapping stars by comparing the positions of all sources and identifying duplicates originating from different images. In cases of duplication, the star with the lower photometric error was retained for simplicity. After this cleaning process, a final mock catalog of all stars across the three simulation regions was produced.

\section{Results}
\label{3}

\subsection{Photometric Result}
\label{3.1}
\begin{figure}
    \centering    
    \includegraphics[height=\textheight, width=0.8\textwidth]{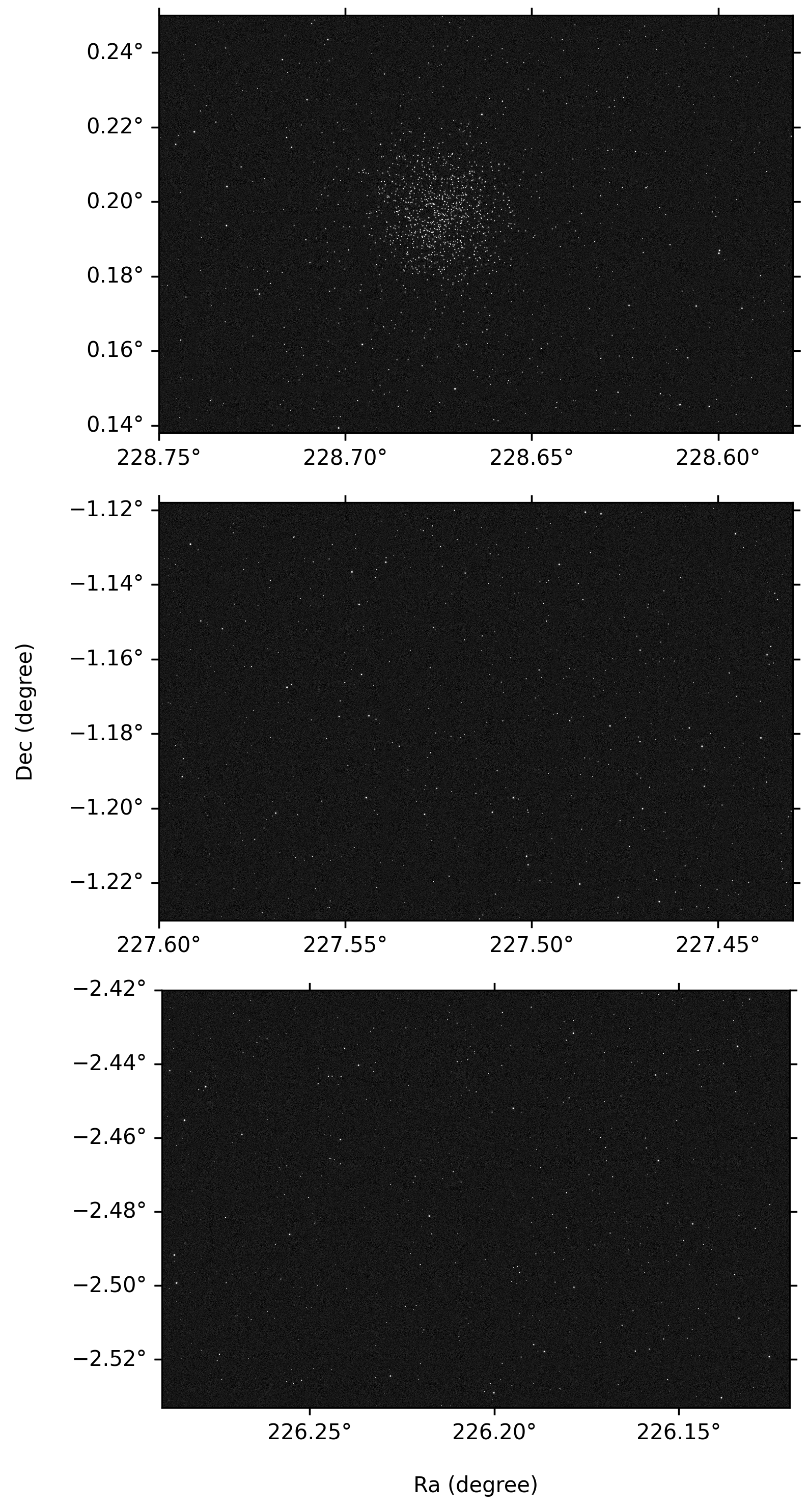}
    \caption{Partial g-band mosaic images of simulated region 1 (upper panel), region 2 (middle panel), and region 3 (lower panel).}
    \label{fig:Figure4}
\end{figure}

The three input star catalogs contain 22,538 (region 1), 10,082 (region 2), and 9,781 (region 3) sources, respectively. To accelerate the image simulation process, the CSST/SC simulation tool employs its built-in filter to pre-select sources based on the saturation magnitude and limiting magnitude of the CSST z-band. Subsequent photometric processing finally produced output catalogs containing 13,226, 5,459, and 5,285 sources for regions 1, 2, and 3, respectively. Figure \ref{fig:Figure4} displays a partial g-band mosaic images for region 1 (upper panel), region 2 (middle panel), and region 3 (lower panel).

\begin{figure}
    \centering    
    \includegraphics[width=\textwidth,height=0.8\textheight,keepaspectratio]{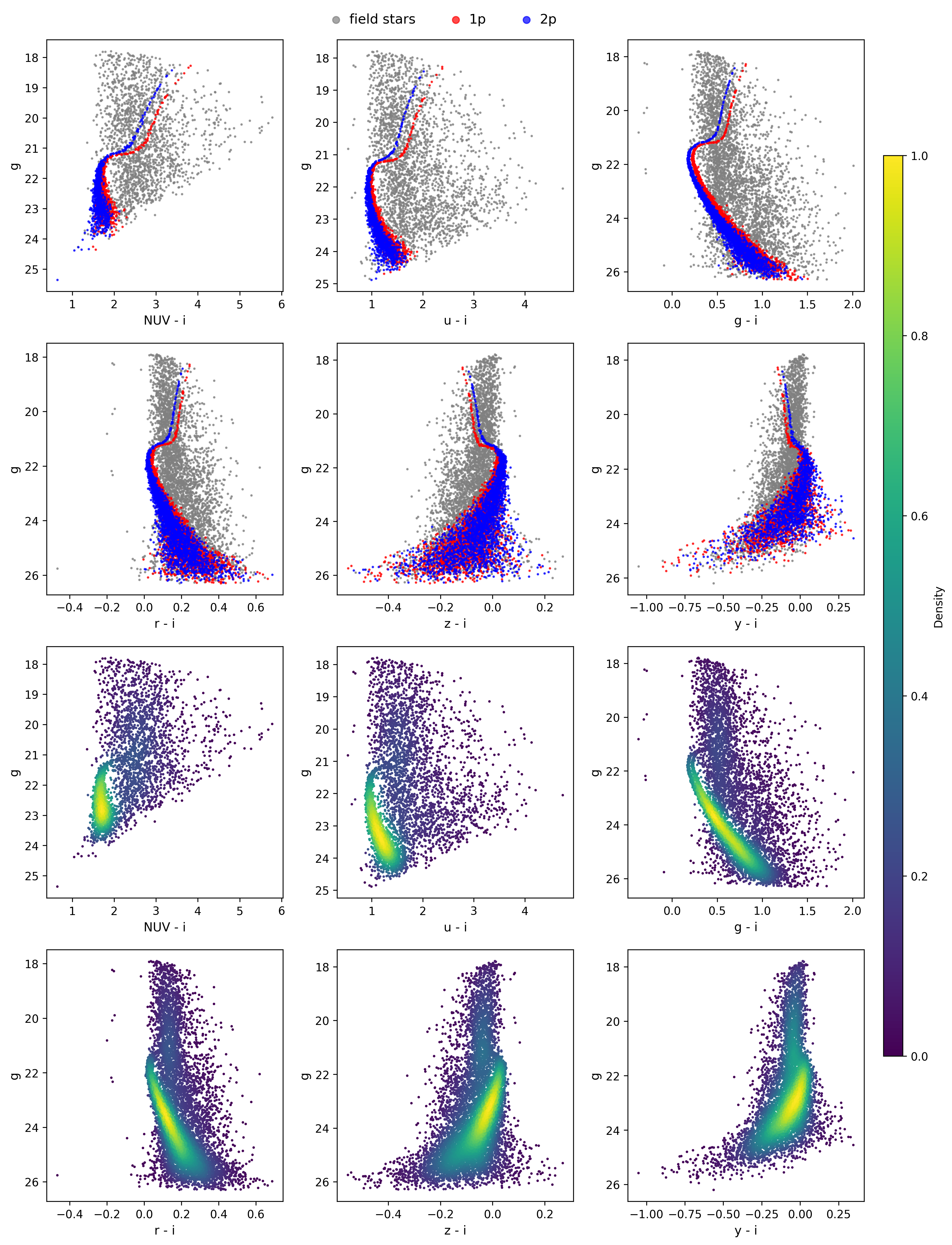}
    \caption{Photometric results for region 1 presented as CMDs. Top two rows: CMDs with stars classified into three populations. Red, blue, and gray dots represent 1P, 2P, and field stars, respectively. Bottom two rows: the same CMDs with stars color-coded by normalized number density (color bar on the right). Panels from left to right and top to bottom in each block show: $g$ vs. $NUV - i$, $u - i$, $g - i$, $r - i$, $z - i$, and $y - i$.}
    \label{fig:Figure5}
\end{figure}

In the study of helium variations in star clusters using CSST/SC, \citet{ji2023study} demonstrated that 2P stars, driven by He-enrichment, are bluer than 1P stars across all color bands. The study further highlighted that the unique CSST/SC $NUV$, $u$ passbands are particularly effective in maximizing the color separation between helium-normal and helium-enriched populations. According to \citet{2015ApJ...808...51M,2015MNRAS.447..927M}, N-enriched stars are also CO-depleted, with key CNO-related spectral features predominantly distributed in the UV to blue wavelength range. Therefore, we expect that the $NUV$ and $u$ bands of CSST/SC will resolve a larger color difference between the 1P and 2P in both MS and RGB phase. Figure \ref{fig:Figure5} presents a series of CMDs in the $g$-band versus various colors to characterize the stellar populations in region 1. We selected the $i$-band as the subtrahend to ensure a long wavelength baseline, thereby emphasizing colors involving UV bands. Data points are color-coded based on known labels (top two rows of Figure \ref{fig:Figure5}): red for 1P, blue for 2P, and gray for field stars. The figure reveals that while CMDs utilizing CSST/SC UV photometry are powerful in disentangling stellar populations on the giant branch, the $g$ vs. $g-i$ CMD provides a clearer distinction between the 1P and 2P on the MS phase at a 150 s exposure. This result appears inconsistent with previous findings by \citet{ji2023study}, who reported diminishing color differences in both the MS and RGB as the minuend filter changes from NUV to i. To further investigate this discrepancy, we turn to bottom two rows of Figure \ref{fig:Figure5}, which presents normalized stellar number density maps constructed without prior knowledge of stellar labels from region 1. These maps corroborate that the $g$-band provides superior separation for the MS phase, whereas on the giant branch, the member stars blend with the field stars, showing no clear separation. Notably, the $NUV$ and $u$ bands fail to reveal clear distinctions between the different populations.

\begin{figure}
    \centering    
    \includegraphics[width=\textwidth]{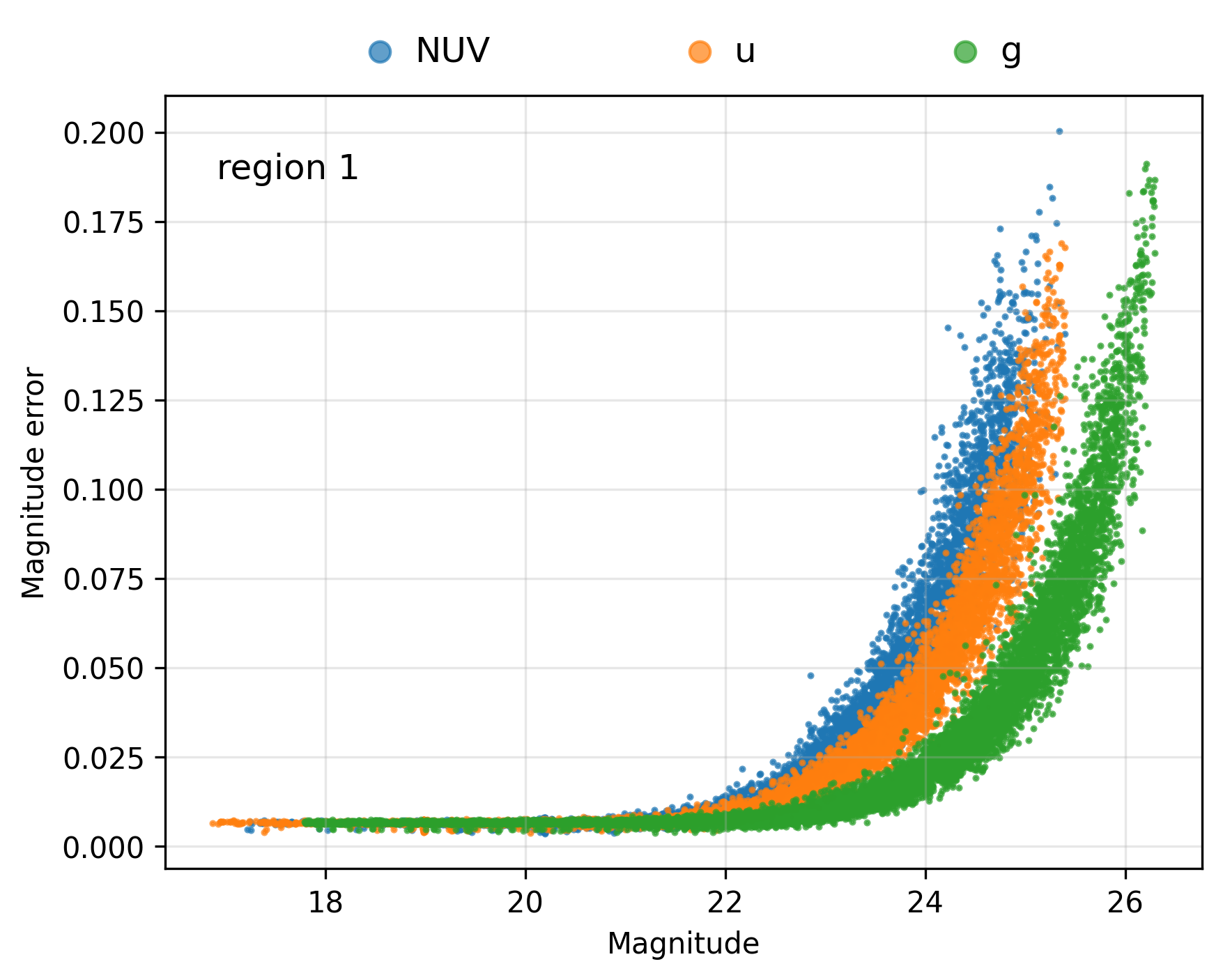}
    \caption{Magnitude error as function of magnitude in region 1 for the $NUV$ (blue points), $u$ (orange points), and $g$ (green) bands.}
    \label{fig:Figure6}
\end{figure}

This phenomenon can be attributed to several factors. First, He-enrichment causes 2P stars to become bluer even in the $g$ band. Second, as reviewed by \citet{Zhan2021}, the system throughput efficiency in the $g$ band is significantly higher than that in the $NUV$ and $u$ bands, allowing better photometric precision. Figure \ref{fig:Figure6} shows that at a comparable magnitude level, the photometric errors in the $g$-band (green points) are systematically and significantly lower than those in the $NUV$ (blue points) and the $u$-band (orange points). This improved precision produces tighter stellar sequences in the CMD, thereby amplifying the subtle color differences between distinct stellar populations and enabling clearer resolution. 

Additionally, the CMDs for region 1 exhibit relatively few field stars in the MS phase compared to the giant phase. This scarcity further challenges the detection of significant color differences in the giant star phase using $NUV$ and $u$ bands during 150-second exposures without prior stellar labels. Regarding the paucity of field stars in the MS phase, we have re-examined the complete Palomar 5 stellar stream dataset and confirmed that this characteristic persists in the region 1 and region 2 (see the APPENDIX). The most probable reason is the dominance of member stars over field stars at an equal distance modulus.

\begin{figure}
    \centering    
    \includegraphics[width=\textwidth,height=0.8\textheight,keepaspectratio]{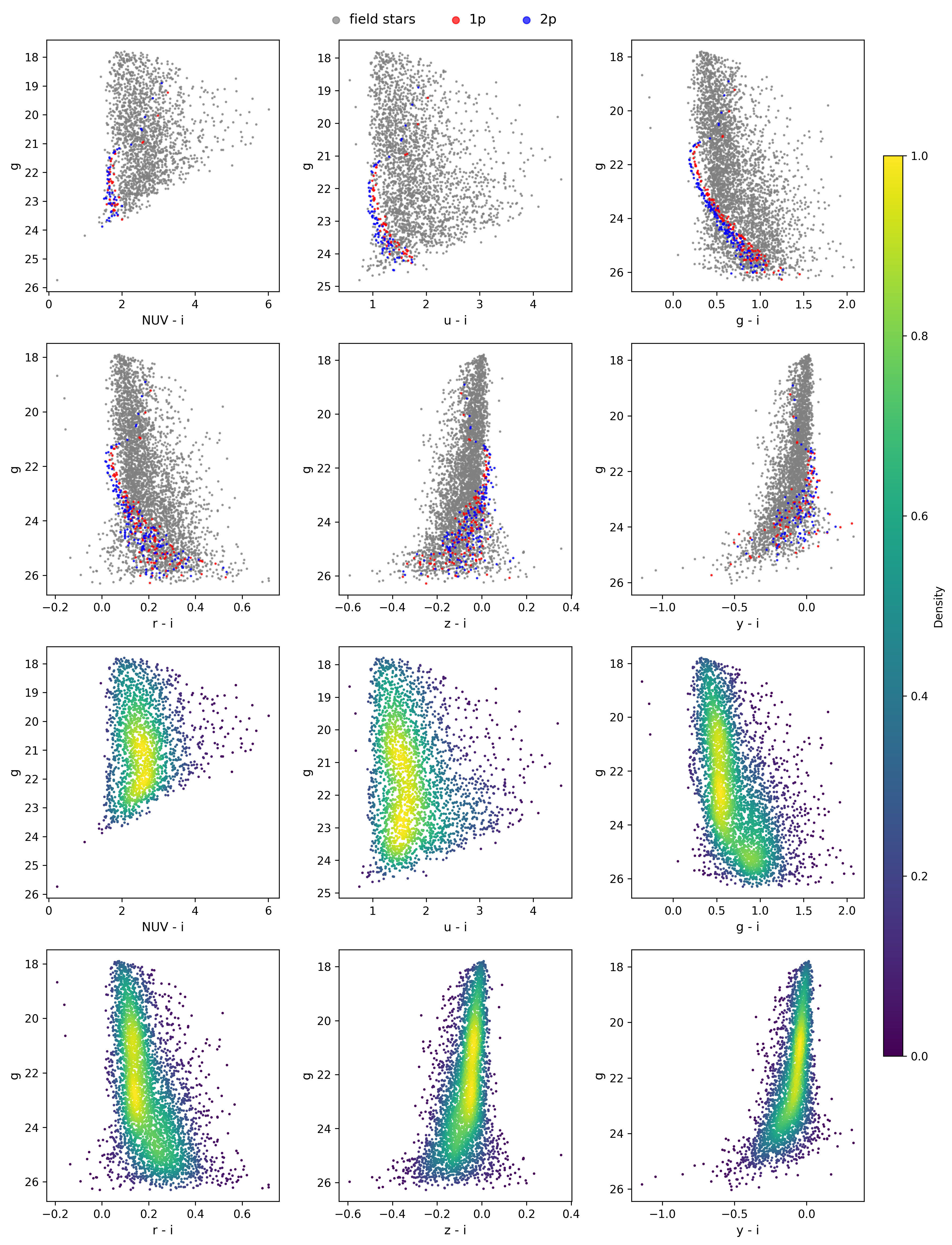}
    \caption{Same as Figure \ref{fig:Figure5}, but for region 2. Top two rows: CMDs with stars classified into three populations. Red, blue, and gray dots represent 1P, 2P, and field stars, respectively. Bottom two rows: the same CMDs with stars color-coded by normalized number density (color bar on the right). Panels from left to right and top to bottom in each block show: $g$ vs. $NUV - i$, $u - i$, $g - i$, $r - i$, $z - i$, and $y - i$.}
    \label{fig:Figure7}
\end{figure}

\begin{figure}
    \centering    
    \includegraphics[width=\textwidth,height=0.8\textheight,keepaspectratio]{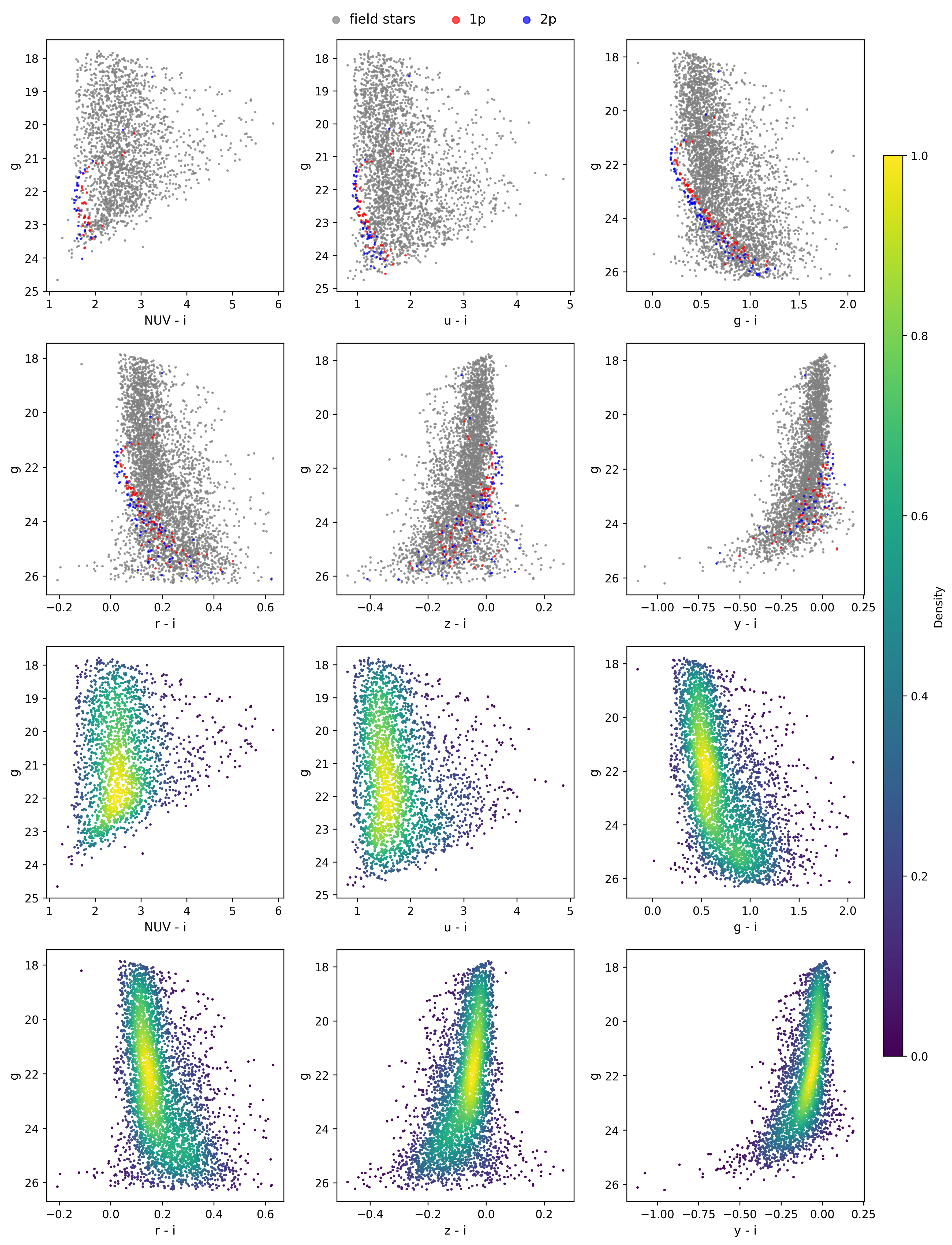}
    \caption{Same as Figure \ref{fig:Figure5}, but for region 3. Top two rows: CMDs with stars classified into three populations. Red, blue, and gray dots represent 1P, 2P, and field stars, respectively. Bottom two rows: the same CMDs with stars color-coded by normalized number density (color bar on the right). Panels from left to right and top to bottom in each block show: $g$ vs. $NUV - i$, $u - i$, $g - i$, $r - i$, $z - i$, and $y - i$.}
    \label{fig:Figure8}
\end{figure}

We performed similar analyses for regions 2 and 3 in Figures \ref{fig:Figure7} and \ref{fig:Figure8}. The photometric precision results of region 2 and region 3 are similar to Figure \ref{fig:Figure6}. The figure is not repeated here. Since regions 2 and 3 are located within the stellar stream of Palomar 5, the number of member stars has significantly decreased. Consequently, even with known stellar types (1P, 2P or field stars), the $NUV$ and $u$ bands can no longer clearly separate the 1P and 2P in the CMDs of Figures \ref{fig:Figure7} and \ref{fig:Figure8}, whereas the $g$ band retains this capability. Furthermore, as observed in Figures \ref{fig:Figure5}, \ref{fig:Figure7}, and \ref{fig:Figure8}, the ability to disentangle stellar populations weakens in the $NUV$, $u$, and $g$ bands as the number of member stars decreases. The same conclusion can be drawn from the CMDs in Figures \ref{fig:Figure7} and \ref{fig:Figure8}, for which no stellar types are provided. 

\subsection{Extent of Stellar Population Contamination}
\label{3.2}
The distance and faintness of the Palomar 5 stream make it particularly challenging to cleanly isolate its member stars from the substantial foreground and background contamination along the line of sight. This is further exacerbated by various factors contributing to the photometric error, such as the photometric noise, unresolved binaries, blending, PSF fitting residuals, etc. These considerations motivate us to evaluate membership assignment in which we combine the CMD information described above with $NUV$, $u$ and $g$ bands photometry. We have presented a rough method for tagging member stars (1P or 2P) and field stars in all three regions. Based on the demonstrated capability of the $NUV$, $u$, and $g$ photometric bands to distinguish MPs (as detailed in Section \ref{3.1}), we constructed theoretical CMDs in the $g$ versus $NUV - i$, $u - i$, and $g - i$ colors using the theoretical apparent magnitudes described in Section \ref{2.4}. We then calculated the identification accuracy of member stars under the judgment of these three theoretical CMDs, respectively. The photometric magnitudes of a star that does not conform to either the 1P or 2P theoretical CMD sequences within photometric error is classified as a field star; otherwise, it is considered a member star. In specific cases where a star is consistent with both sequences, its classification (1P or 2P) is determined by comparing its distance to each fiducial sequence. Table \ref{tab:2} presents the classification results of member stars and field stars for the three regions. In the parentheses in Table \ref{tab:2}, we include the contamination rate and completeness rate for each category, defined as:

\begin{equation}
    \text{contamination rate} = \frac{\text{number of samples misclassified into this category}}{\text{total number of samples classified into this category}},
\end{equation}

and

\begin{equation}
    \text{completeness rate} = \frac{\text{number of correctly classified samples}}{\text{total number of samples that truly belong to this category}}.
\end{equation}

\begin{table}
\centering
\renewcommand\tabcolsep{12pt}
\caption{Classification results for different regions and color indexes. The first number in parentheses indicates the contamination rate, and the second number indicates the completeness rate. Notably, the classification results were obtained by integrating photometric magnitudes derived from 150 s exposure simulated images with theoretical CMDs.}
\begin{tabular}{lccc}
\hline
\hline
& Region 1 & Region 2 & Region 3 \\
\hline
\textbf{Initial classification} &  \\
\cline{1-1}
Total stars & 13226 & 5459 & 5285 \\
Field stars & 4996 & 4898 & 4976 \\
Member stars  & 8230 & 561 & 309 \\
\quad 1P stars & 3982 & 234 & 142 \\
\quad 2P stars & 4248 & 327 & 167 \\
\hline
\textbf{After $NUV - i$ classification} & \\
\cline{1-1}
Total stars & 13226 & 5459 & 5285 \\
Field stars & 13051 (61.72\%, 100\%) & 5449 (10.11\%, 100\%) & 5278 (5.72\%, 100\%) \\
Member stars & 175 (0\%, 2.13\%) & 10 (0\%, 1.78\%) & 7 (0\%, 2.27\%) \\
\quad 1P stars & 85 (0\%, 2.13\%) & 2 (0\%, 0.85\%) & 3 (0\%, 2.11\%) \\
\quad 2P stars & 90 (0\%, 2.12\%) & 8 (0\%, 2.45\%) & 4 (0\%, 2.40\%) \\
\hline
\textbf{After $u - i$ classification} &  \\
\cline{1-1}
Total stars & 13226 & 5459 & 5285 \\
Field stars & 12332 (59.50\%, 99.96\%) & 5397 (9.25\%, 100\%) & 5247 (5.16\%, 100\%) \\
Member stars & 894 (0.22\%, 10.84\%) & 62 (0\%, 11.05\%) & 38 (0\%, 12.3\%) \\
\quad 1P stars & 460 (0\%, 11.55\%) & 26 (0\%, 11.11\%) & 18 (0\%, 12.68\%) \\
\quad 2P stars & 434 (0.46\%, 10.17\%) & 36 (0\%, 11.01\%) & 20 (0\%, 11.98\%) \\
\hline
\textbf{After $g - i$ classification} & \\
\cline{1-1}
Total stars & 13226 & 5459 & 5285 \\
Field stars & 10632 (53.10\%, 99.80\%) & 5306 (7.88\%, 99.80\%) & 5194 (4.39\%, 99.80\%) \\
Member stars & 2594 (0.39\%, 31.40\%) & 153 (6.54\%, 25.49\%) & 91 (10.99\%, 26.21\%) \\
\quad 1P stars & 1349 (0.30\%, 33.78\%) & 67 (8.96\%, 26.07\%) & 48 (14.58\%, 28.87\%) \\
\quad 2P stars & 1245 (0.48\%, 29.17\%) & 86 (4.65\%, 25.08\%) & 43 (6.98\%, 23.95\%) \\
\hline
\end{tabular}
\label{tab:2}
\end{table}

From Table \ref{tab:2}, classifications using the $NUV-i$ color achieve near-perfect purity (contamination rate $\approx 0\%$ for member stars in all regions), but at the cost of very low completeness ($\sim2\%$). This suggests that while the few stars selected by $NUV-i$ are highly reliable members, the $NUV$ band's faintness and associated large photometric errors cause the vast majority of genuine members to be missed. In contrast, the $g-i$ color offers a more balanced performance. It maintains excellent purity (contamination rate $<1\%$ in region 1) while achieving significantly higher completeness ($\sim31\%$ in region 1), making it the most effective single color for member identification. The $u-i$ color performs intermediately. Furthermore, the efficacy of all color indices deteriorates in stream regions (regions 2 and 3), where the field-to-member ratio is extreme. For instance, the contamination rate for ($g-i$)-selected members in region 3 rises to $\sim11\%$. This indicates that in the sparse stellar stream, CMDs alone are insufficient for reliable member discrimination due to the overwhelming field star contamination. Consequently, robust member identification in these regions will necessitate the integration of additional parameters, such as deep photometry, proper motions, and radial velocities.

\subsection{Deep Photometry for Palomar 5}
\label{3.3}

CSST will conduct a 10-year survey, performing multi-color photometric imaging and slitless spectroscopy over 17,500 $\deg^2$ in wide-field and 400 $\deg^2$ in deep-field modes. The CSST/SC is responsible for these primary observational tasks. For the wide-field survey, each sky area will be covered at least twice in each photometric band, with a single exposure of $\approx$150 s. To detect MPs of MS and RGB phases in the $g$ band, and to ensure a SNR of no less than 100 in the $NUV$ and $u$ bands, we used the online CSST exposure time calculator (ETC) to estimate that the required exposure time needs to be at least 150s $\times$ 10 (1500 s). Therefore, we stacked ten 150 s exposures from three regions using SWarp \citep{2002ASPC..281..228B}. This stacking procedure will allow us to assess the CSST's ability to resolve different stellar populations under deep photometric conditions. Based on the previous analysis, we only discuss the deep photometry of the $NUV$, $u$, and $g$ bands.

\begin{figure}
    \centering        \includegraphics[width=\textwidth,height=0.8\textheight,keepaspectratio]{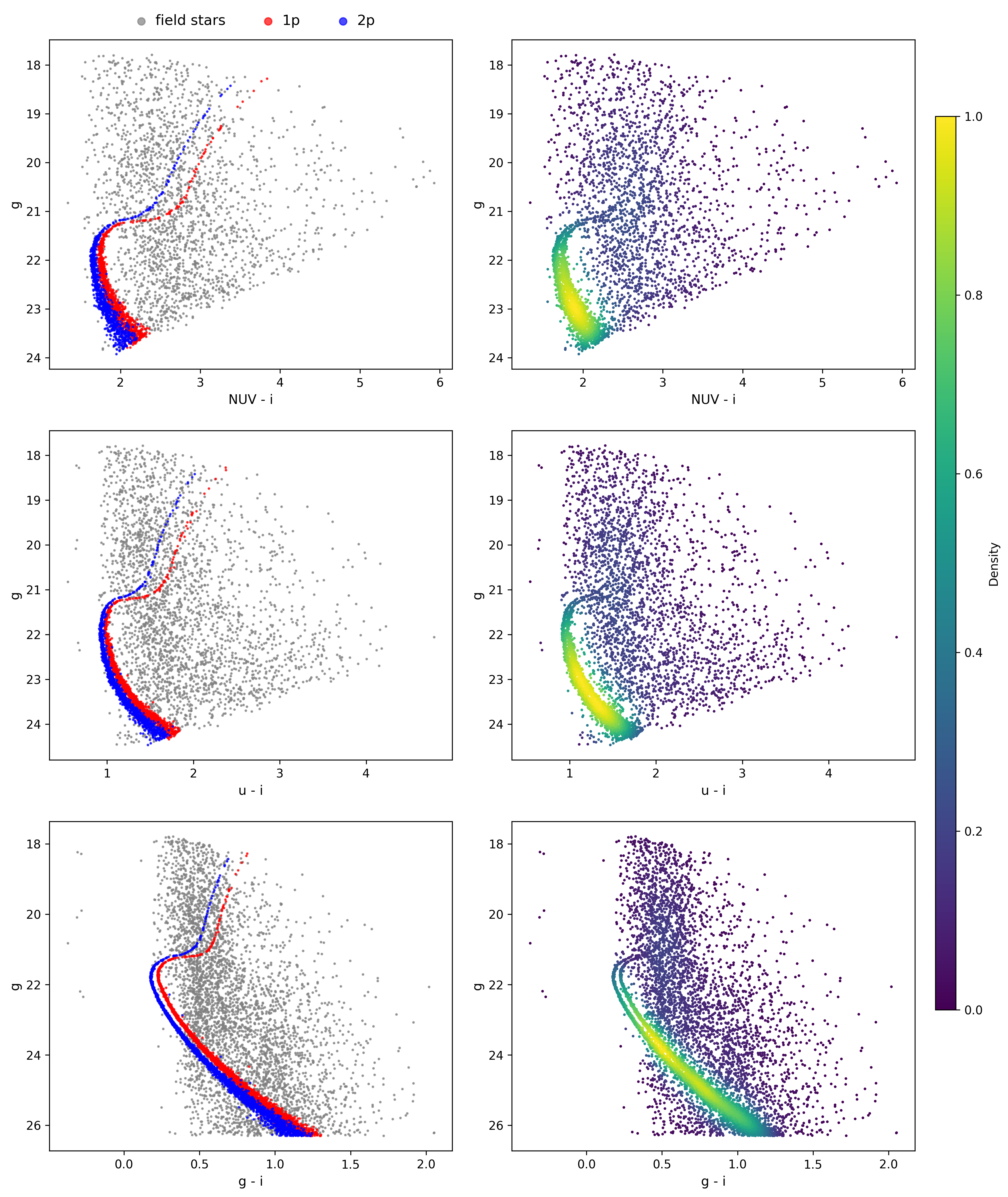}
    \caption{The deep photometric results (150 s $\times$ 10) of region 1 in the form of CMDs. Left column: stars are categorized by known labels. Right column: stars are color-coded by normalized number density. From top to bottom panels: $g$ vs. $NUV - i$, $u - i$, $g - i$.}
    \label{fig:Figure9}
\end{figure}

\begin{figure}
    \centering        \includegraphics[width=\textwidth,height=0.8\textheight,keepaspectratio]{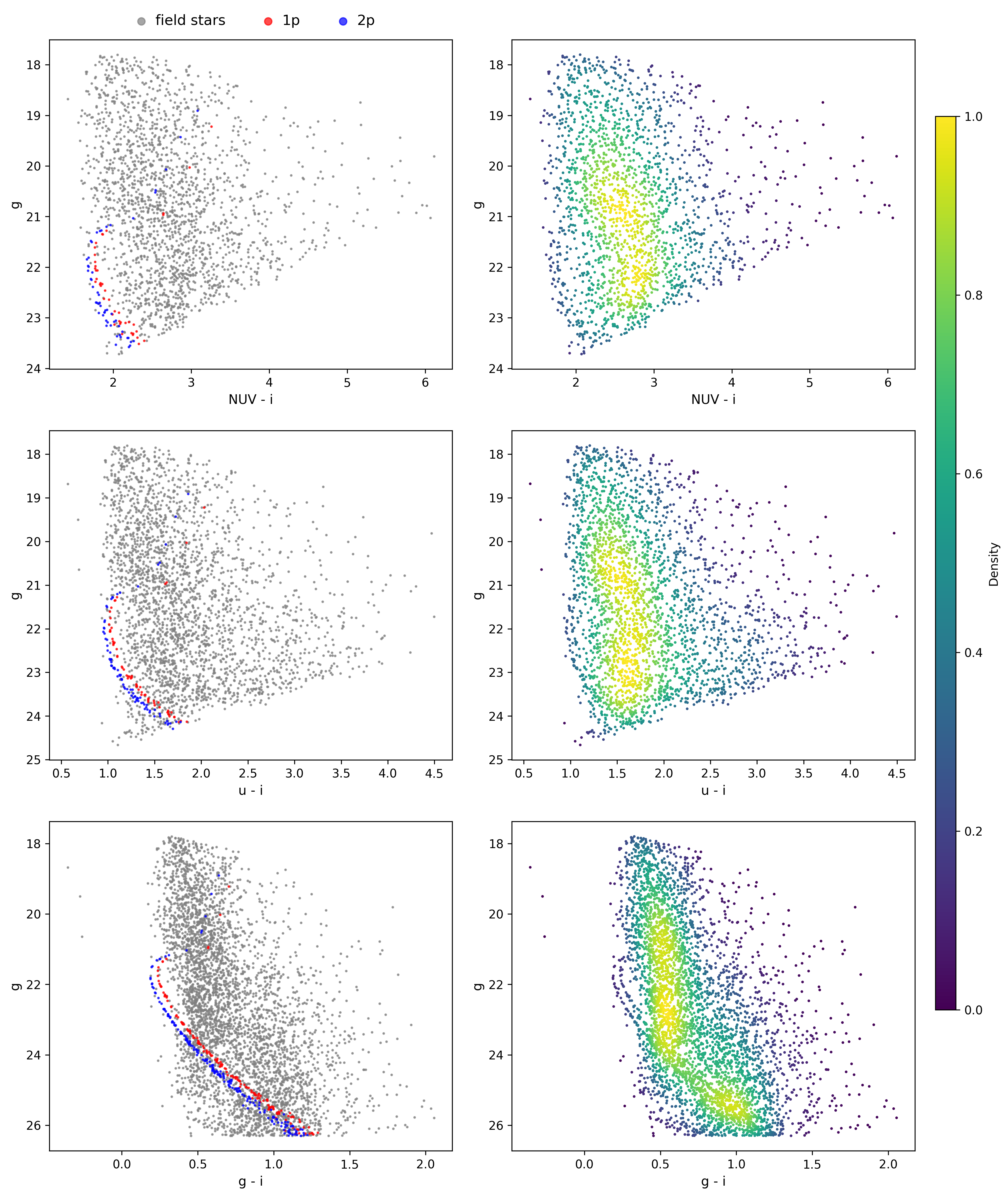}
    \caption{The deep photometric results (150 s $\times$ 10) of region 2 in the form of CMDs. Left column: stars are categorized by known labels. Right column: stars are color-coded by normalized number density. From top to bottom panels: $g$ vs. $NUV - i$, $u - i$, $g - i$.}
    \label{fig:Figure10}
\end{figure}

\begin{figure}
    \centering        \includegraphics[width=\textwidth,height=0.8\textheight,keepaspectratio]{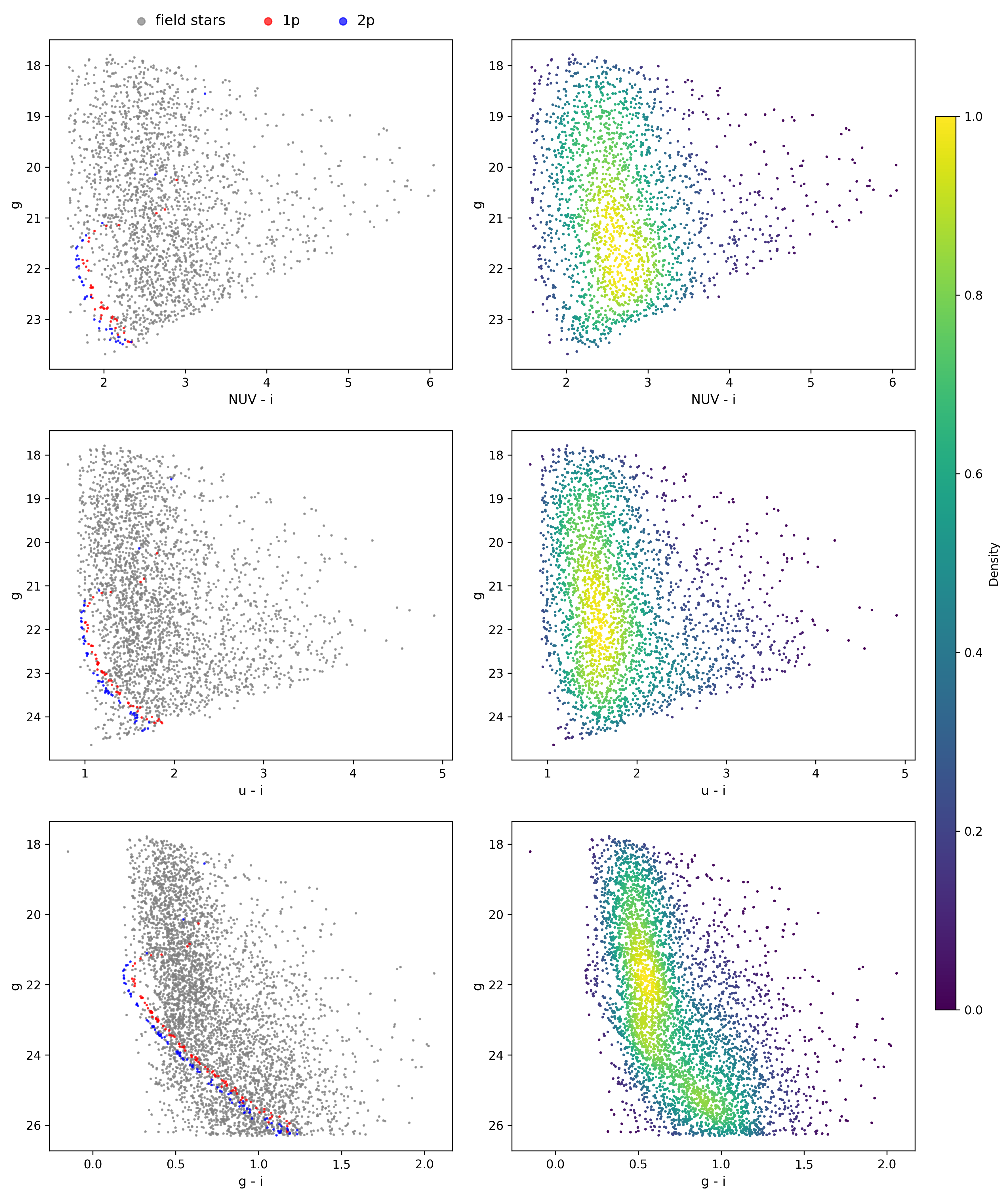}
    \caption{The deep photometric results (150 s $\times$ 10) of region 3 in the form of CMDs. Left column: stars are categorized by known labels. Right column: stars are color-coded by normalized number density. From top to bottom panels: $g$ vs. $NUV - i$, $u - i$, $g - i$.}
    \label{fig:Figure11}
\end{figure}

Figures \ref{fig:Figure9}, \ref{fig:Figure10}, and \ref{fig:Figure11} present the results of deep photometry for regions 1, 2, and 3, respectively. With the exception of the sparse regions of the stream, the left panels demonstrate that, if member stars can be robustly identified, deep photometry in all three bands can, in principle, distinguish MPs in both the MS and RGB phases. The $g$-band reveals a more pronounced separation between the 1P and 2P MS phase compared to a single exposure of 150 s. In the cases of unknown classifications, we found that even after stacking ten exposures in the $NUV$ band (which improves the SNR by approximately a factor of 3.16), no clear distinction is observed between the different stellar populations on the MS. Furthermore, the stacked data in all bands fail to reveal MPs in the RGB phase. 

The difficulty in resolving MPs in extended streams like Palomar 5 directly from CMDs underscores the necessity of proper motion data. While photometric surveys from the HST and the upcoming CSST/SC provide valuable color-magnitude information, precise kinematic measurements remain essential to disentangle stream members from foreground and background contamination. \citet{starkman2020extended} used proper motion data from Gaia data release 2 (Gaia DR2) to select members of the Palomar 5 stream. Their analysis showed that the proper motion distribution of stars within two projected tidal radii of Palomar 5 is sharply peaked around -2.7 $\pm$ 0.1 (mas yr$^{-1}$) \citep{starkman2020extended}. Based on this peak, they adopted a circular region with a radius of 3 mas yr$^{-1}$ centered at ($\mu_{\alpha^{*}}$, $\mu_{\delta}$) $=$ (-2.5, -2.5) mas yr$^{-1}$ as a proper motion selection window. This window efficiently removed 81.9\% of field stars. However, due to the distance of Palomar 5, its main sequence turn-off stars lie near Gaia's detection limit ($G$ $\approx$ 21 mag). Even with the improved precision of Gaia data release 3 (Gaia DR3), the bulk of the stream’s stellar population remains not well resolved in Gaia data \citep{2025NewAR.10001713B}. In Gaia DR2, the proper motion uncertainty is approximately 0.07, 0.2, 1.2, and 3 mas yr$^{-1}$ at $G$ $<$ 15 mag, $G$ $=$ 17, 20, and 21 mag, respectively \citep{2018A&A...616A...1G}. CSST will also possess a strong astrometric capability. Under ideal conditions, CSST is expected to achieve proper motion accuracies $<$ 1 mas yr$^{-1}$ for stars between 18-22 mag in $g$ band, and $>$ 1 mas yr$^{-1}$ for those at 22-26 mag in $g$ band \citep{fu2023simulation}. By combining proper motion data from Gaia and CSST, we can identify member stars in both the MS and RGB phases. Once CSST is fully operational, we plan to observe the Palomar 5 cluster center and both high- and low-density regions of its tidal tails. By conducting observations across ten different epochs in multiple bands ($NUV$, $u$, $g$, $i$), we aim to obtain the deep photometry and proper motion data necessary for a definitive MP analysis.

Finally, based on Figure \ref{fig:Figure6}, a rough estimate suggests that to achieve comparable photometric precision in $NUV$ band as in the $g$-band for detecting MPs under 150 s exposures, the magnitude difference is approximately 2. Using the formula $m_1-m_2 =  5\log_{10}{\left ( \frac{d_1}{d_2} \right )}$, we estimate that the observed stream would need to be located at a distance of $\lesssim$ 8 kpc.

\section{Discussion}
\label{4}


The investigation into the chemical abundance of Palomar 5 dates back to the initial spectroscopic study by \citet{smith1985spectroscopya}, who identified a distinct CH–CN anticorrelation in the cluster. Their work demonstrated that this characteristic chemical signature could be detected even within a relatively small sample of giant stars. This finding was later extended by \citet{smith2002study} in the first high-resolution spectroscopic study of four Palomar 5 giants with high SNR measurements, which revealed significant abundance variations in light elements such as C, N, Na, and Al. By co-adding high-resolution spectra of 15 giant stars from the main body of Palomar 5, \citet{koch2017galactic} concluded that this method remains insensitive to chemical spreads. More recently, using data from the Apache Point Observatory Galactic Evolution Experiment (APOGEE), \citet{phillips2022apogee} analyzed the chemical compositions of seven RGB and AGB stars located in both the cluster body and the tidal streams of Palomar 5. The study categorized these stars into N-normal and N-rich groups, noting that the N-rich stars did not exhibit corresponding Al enrichment and showed only a minor spread in [O/Fe]. 

Based on the above research, Palomar 5 exhibits the typical light element abundance anomalies (such as variations in C, N, O, and Na) characteristic of GCs. However, its chemical evolution pattern may be unique, as highlighted by the fact that its N-rich stars are not accompanied by significant Al enrichment. Theoretically, He, C, N, O, Na, Mg, and Al are the most abundant elements in stellar atmospheres, producing strong absorption features. However, as demonstrated by \citet{2022RAA....22i5004L}, the selected CSST filters are ineffective in identifying MPs with varying abundances of Na, Mg, and Al, as the associated photometric effects are negligible or too small to be feasibly detected. Therefore, we only consider the variations of He, C, N, O in this work. Also according to the work of \citet{2022RAA....22i5004L}, the presence of CNO variations will nullify the color difference between stellar populations arising from He enrichment alone. Observations of local globular clusters show that the estimated helium content for the 2P could decrease from Y $\simeq$ 0.4 to as low as Y $\simeq$ 0.3 \citep{2007ApJ...661L..53P,portinari2010revisiting,2018MNRAS.481.5098M,2025arXiv250717057B}. Currently, the helium mass fraction $Y$ of Palomar 5 has not been directly measured. For the 2P isochrone in Section \ref{2.2}, we adopt a helium mass fraction of Y=0.33 because the Dartmouth Stellar Evolution Database only provides the enriched-Helium mass fraction of 0.33 and 0.40. And Y=0.40 is a suitable stellar model to describe an NGC 2808-like GC with an extreme helium enrichment. Consequently, future mock observations can systematically test different chemical patterns, such as models with isolated He or CNO variations and a broader range of helium abundances.

\section{Conclusion}
\label{5}

In this work, we performed mock observations of the stellar stream Palomar 5 for CSST/SC, while considering the effects of field stars, photometric noise, PSF, and variations in He, C, N, and O abundances. Our primary goal was to test and validate an observational methodology for studying MPs in stellar streams with the CSST/SC, thereby preparing for future efficient CSST observations.

Our simulations under 150 s exposures identify the $g$-band as the optimal filter for distinguishing 1P and 2P populations on the MS, a direct result of its superior photometric precision. Although the $NUV$ and $u$-bands cover critical spectral features, their lower photometric accuracy renders them impractical for resolving the MPs. For RGB stars, population separation remains challenging across all bands due to blending with field stars, the issue exacerbated in sparse stream regions by declining member star counts. We further evaluated member star identification using theoretical CMDs, revealing that the CSST/SC $g$-band provides the optimal balance between contamination rate and completeness rate in Palomar 5's core. However, this band fails in the stream due to severe field star contamination, unequivocally highlighting the necessity of other methods for robust membership determination.

To improve the SNR, we simulated ten exposures of 150 s for all three regions and subsequently stacked the simulation images. Even after stacking $NUV$ band images to improve the SNR, distinct separation between different stellar populations on the MS remains unachievable. Furthermore, the stacked data in all bands failed to detect any MP signals among the RGB. Our results highlight that even with CSST’s high UV sensitivity and large FoV, the detection of MPs in distant, extended streams is limited in the absence of proper motion data. Future CSST observations of Palomar 5 and its tidal tails will employ ten different epochs in multiple bands ($NUV$, $u$, $g$, $i$) to provide the deep photometry (26 mag in $g$ band) and proper motions (uncertainty: $<$ 1 mas yr$^{-1}$ at $g$ = 18-22 mag and $>$ 1 mas yr$^{-1}$ at $g$ = 22-26 mag) necessary for a definitive MP analysis. 


Although actual CSST observations will also be affected by factors such as extinction and complex instrumental effects, this work establishes a framework and offers practical guidance for planning future CSST/SC observations. Future studies will focus on nearer stellar streams ($\lesssim$ 8 kpc) to further explore the potential of the telescope in resolving MPs. 

\begin{acknowledgments}
We thanks the support from the National Natural Science Foundation of China through grant 12573041 and 12233013, the High-level Youth Talent Project (Provincial Financial Allocation) through the grant 2023HYSPT0706, the one-hundred-talent project of Sun Yat-sen University, the Fundamental Research Funds for the Central Universities, Sun Yat-sen University (2025QNPY04), and the support from the China Manned Space Project with NO.CMS-CSST-2021-A08.

\end{acknowledgments}

\begin{contribution}

Xia Li contributed to method development, data analysis, and manuscript writing.
Long Wang camp up with the initial research concept, provided simulation data, contributed to discussions, edited the manuscript, and supervised the overall project. Chengyuan Li instructed on the use of the \textsc{spectrum} tool and contributed to discussions. Yang Chen advised on the application of the \textsc{YBC} bolometric corrections database. Hao Tian generated the field star catalog using the Galaxia code. Xin Zhang guided the use of the CSST mock observation code.


\end{contribution}

%
\facilities{CSST}

\software{astropy \citep{2013A&A...558A..33A,2018AJ....156..123A,2022ApJ...935..167A},
NumPy \citep{harris2020array},
matplotlib \citep{2007CSE.....9...90H},
\textsc{petar} \citep{wang2020PETAR} \url{https://github.com/lwang-astro/PeTar},
\textsc{sdar} \citep{2020MNRAS.493.3398W};\url{https://github.com/lwang-astro/SDAR},
\textsc{sse}/\textsc{bse} \citep{hurley2000comprehensive, hurley2002evolution, banerjee2020bse},
\textsc{galpy} \citep{bovy2015galpy},
Dartmouth Stellar Evolution \citep{dotterDartmouthStellarEvolution2008};\url{https://rcweb.dartmouth.edu/stellar/},
\textsc{spectrum} \url{https://www.appstate.edu/~grayro/spectrum/spectrum.html},
\textsc{ybc} bolometric corrections database \citep{2019A&A...632A.105C};\url{https://sec.center/YBC/index.html},
CSST mock observation code; \url{https://csst-tb.bao.ac.cn/code/csst-sims/csst_msc_sim},
Galaxia code \citep{2011ApJ...730....3S}; \url{https://galaxia.sourceforge.net/Galaxia3pub.html},
Photutils \citep{larry_bradley_2024_13989456};
\url{https://photutils.readthedocs.io/en/stable/},
the online CSST exposure time calculator; \url{https://nadc.china-vo.org/csst-planning/tools/etc_msc},
SWarp \citep{2002ASPC..281..228B}; \url{https://www.astromatic.net/software/swarp/}
}

\appendix

\section{HR Diagram and Distance Modulus Distribution}
\begin{figure}
    \centering    
    \includegraphics[width=0.8\textwidth]{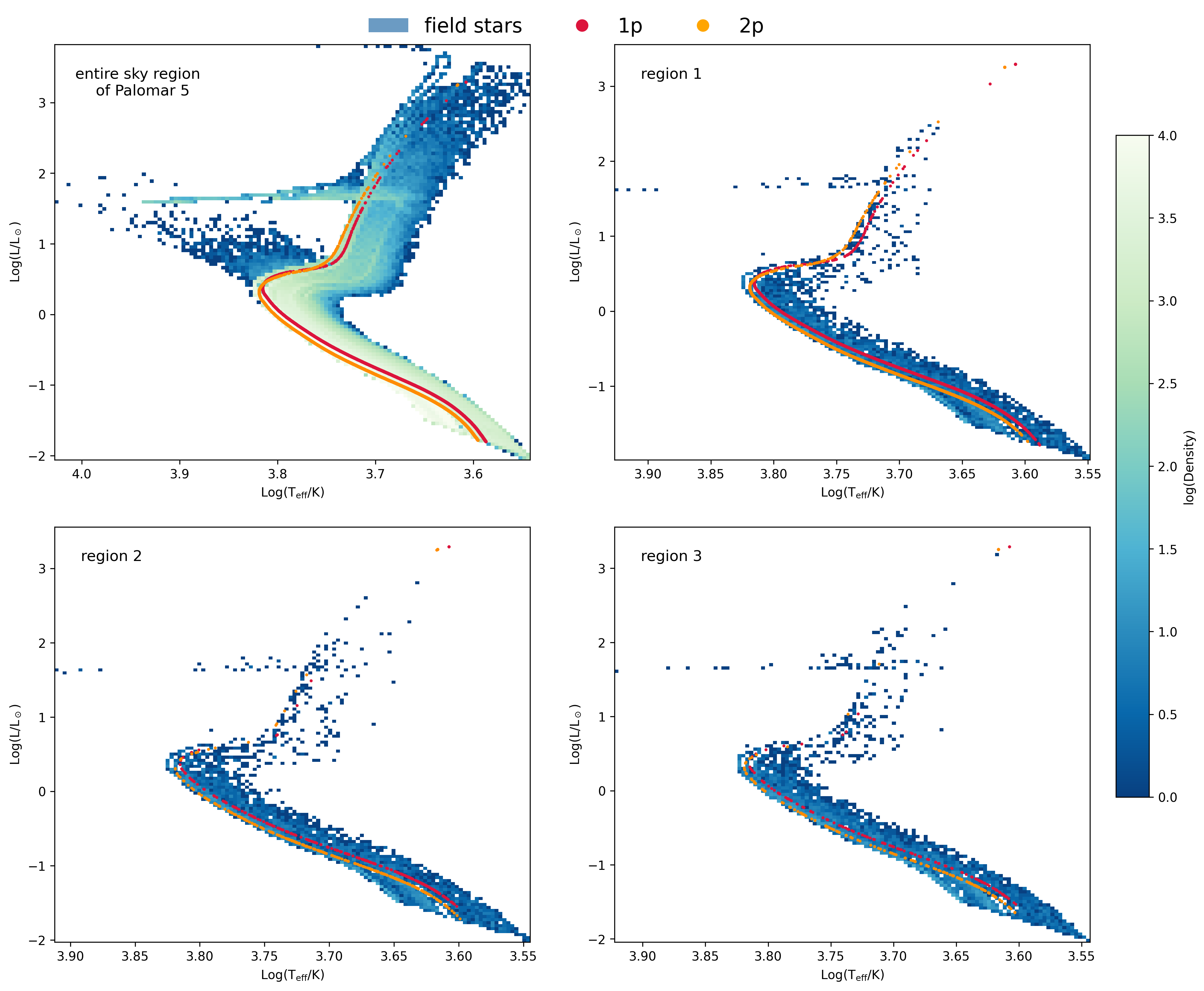}
    \caption{Theoretical isochrones for Palomar 5 are displayed in the plane of $\log{(L/L_\odot})$ versus $\log{(T_{\mathrm{eff}})}$. Red and organge dots denote 1P and 2P stars, respectively. Field stars are  are color-coded (GnBu colormap) by log(density). From left to right and top to bottom: entire sky region of Palomar 5, region 1, region 2, and region 3.}
    \label{fig:Figure12}
\end{figure}

\begin{figure}
    \centering    
    \includegraphics[width=0.8\textwidth]{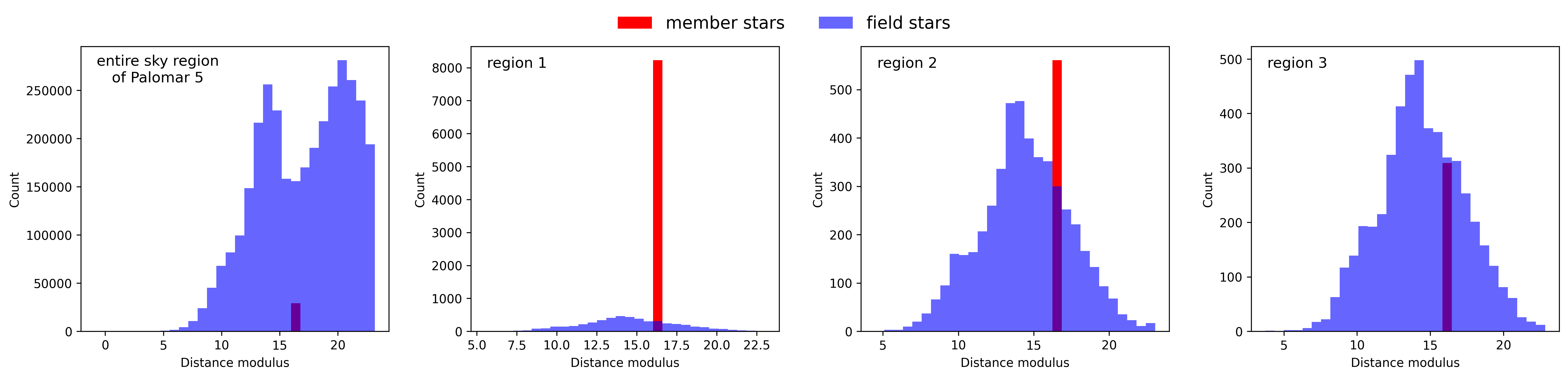}
    \caption{The distribution of distance modulus. Red bars are for member stars and blue ones for the field stars. From left to right: entire sky region of Palomar 5, region 1, region 2, and region 3.}
    \label{fig:Figure13}
\end{figure}

Figure \ref{fig:Figure12} presents the isochrones for the entire sky region and different areas of Palomar 5, while Figure \ref{fig:Figure13} shows the distribution of distance modulus across these regions. At the same distance modulus, the number of member stars is significantly larger than that of field stars in the cluster center (region 1). This number decreases in the dense stream area (region 2) but remains higher than the field star count. In the less dense stream area (region 3), however, the number of member stars becomes nearly equal to that of the field stars.

\bibliography{sample701}{}
\bibliographystyle{aasjournalv7}


\end{document}